\documentclass[aps,prl,twocolumn,superscriptaddress,amsfont,graphicx,nofootinbib,preprintnumbers]{revtex4}%
\usepackage{color,graphicx,epsfig}
\usepackage{ifpdf}
\usepackage{amsmath}
\usepackage{bm}
\usepackage{color}
\usepackage[english]{babel}
\usepackage{graphicx}%
\usepackage{amsfonts}%
\usepackage{amssymb}
\usepackage{braket}
\usepackage{hyperref}

\definecolor{nicered}{rgb}{0.7,0.1,0.1}
\definecolor{nicegreen}{rgb}{0.1,0.5,0.1}
\hypersetup{colorlinks,citecolor= nicegreen,linkcolor= nicered}

\newcommand{\slashed}{\slash \hspace{-0.19cm}}

\newcommand{\beq}{\begin{equation}}
\newcommand{\eeq}{\end{equation}}
\newcommand{\bea}{\begin{eqnarray}}
\newcommand{\eea}{\end{eqnarray}}

\definecolor{Red}{rgb}{1.,0.,0.}

\def\mysection#1{{{\bf #1}.~}}
\begin{document}

\def\LjubljanaFMF{Faculty of Mathematics and Physics, University of Ljubljana,
 Jadranska 19, 1000 Ljubljana, Slovenia }
\def\Cincy{Department of Physics, University of Cincinnati, Cincinnati, Ohio 45221,USA}
\def\LjubljanaIJS{Jo\v zef Stefan Institute, Jamova 39, 1000 Ljubljana, Slovenia}

\title{Implications of lepton flavor universality violations in $B$ decays}

\author{Svjetlana Fajfer} 
%\email[Electronic address:]{svjetlana.fajfer@ijs.si} 
\affiliation{\LjubljanaIJS}
\affiliation{\LjubljanaFMF}

\author{Jernej F.\ Kamenik} 
%\email[Electronic address:]{jernej.kamenik@ijs.si} 
\affiliation{\LjubljanaIJS}
\affiliation{\LjubljanaFMF}

\author{Ivan Ni\v sand\v zi\'c} 
%\email[Electronic address:]{ivan.nisandzic@ijs.si} 
\affiliation{\LjubljanaIJS}

\author{Jure Zupan} 
%\email[Electronic address:]{zupanje@ucmail.uc.edu} 
\affiliation{\Cincy}

\date{\today}
\begin{abstract}
Present measurements of $b\to c\tau\nu$ and $b\to u\tau \nu$ transitions differ from the standard model predictions of lepton flavor universality by almost $4\sigma$. We examine new physics interpretations of this anomaly. An effective field theory analysis shows that minimal flavor violating models are not preferred as an explanation, but are also not yet excluded. Allowing for general flavor violation, right-right vector and right-left scalar quark currents are identified as viable candidates. We discuss explicit examples of two Higgs doublet models,
leptoquarks  as well as quark and lepton compositeness. Finally, implications for LHC searches and future measurements at the (super)B-factories are presented. 
\end{abstract}

\maketitle

%%%%%%%%%%%%%%%%%%%%%%%%%%%%%%
\mysection{Introduction}
%%%%%%%%%%%%%%%%%%%%%%%%%%%%%%
Universality of weak interactions is one of the key predictions of the standard model (SM).
The BaBar collaboration recently performed a test of its consequences in semileptonic $b\to c$ quark transitions via measurements of $B\to D^{(*)}\tau \nu$ branching fractions normalized to the corresponding $B\to D^{(*)}\ell \nu$ modes (with $\ell=e,\mu$)~\cite{:2012xj}
\bea
\mathcal R^*_{\tau/\ell} &\equiv \frac{\mathcal B(B\to D^* \tau \nu)}{\mathcal B (B\to D^{*}\ell \nu)} = 0.332 \pm 0.030\,, \label{eq:Rstar}\\
\mathcal R_{\tau/\ell} &\equiv \frac{\mathcal B(B\to D \tau \nu)}{\mathcal B(B\to D^{}\ell \nu)} = 0.440\pm 0.072\,,  \label{eq:R}
\eea
where the statistical and systematic uncertainties have been combined in quadrature.
%The two ratios, $R^*_{\tau/\ell}$ and $R_{\tau/\ell}$, are excellent probes of new physics (NP), since the dependence of the SM predictions on the hadronic form factors cancels to a large extent. 
Both values in Eqs.~\eqref{eq:Rstar}, \eqref{eq:R} are consistent with previous measurements~\cite{BDTold}, but are also significantly larger  (at $3.4\sigma$ significance when combined) than the SM values $\mathcal R_{\tau/\ell} ^{*,\rm SM} = 0.252(3)$ and $\mathcal R_{\tau/\ell} ^{\rm SM} = 0.296(16)$~\cite{BDTSM}.  If confirmed, this would signal a violation of lepton flavor universality (LFU) in semileptonic $b\to c$ 
transitions at the ${\mathcal O}(30\%)$ level. 

Intriguingly, there are also hints of LFU violations in semileptonic $b\to u $ transitions. The most recent world average of the leptonic $B\to \tau \nu$ branching fraction measurements
$\mathcal B(B^-\to \tau^- \bar \nu) = (11.4\pm2.3)\times 10^{-5}$\cite{BT},
is somewhat larger than its SM prediction with $V_{ub}$ CKM element taken from the global fit~\cite{Charles:2011va}. In contrast, the measured exclusive semileptonic $b\to u\ell\nu$ transition branching fraction
$\mathcal B(\bar B^0\to \pi^+ \ell^- \bar \nu) = (14.6\pm0.7)\times 10^{-5}$~\cite{BPT,LFUPK} 
is consistent with the CKM unitarity predictions~\cite{Laiho:2012ss}. One can get rid of $V_{ub}$ dependence by considering the ratio 
\beq
\mathcal R^{\pi}_{\tau/\ell} \equiv \frac{\tau(B^0)}{\tau(B^-)}\frac{\mathcal B(B^-\to \tau^-\bar\nu)}{\mathcal B(\bar B^0 \to \pi^+\ell^-\bar \nu)} = 0.73\pm0.15\,.
\label{eq:Bpitaunu}
\eeq 
The SM prediction is $\mathcal R^{\pi, \rm SM}_{\tau/\ell} = 0.31(6)$, where we have used the recent Lattice QCD estimates of the relevant $B\to\pi$ form factor and the $B$ decay constant~\cite{Laiho:2009eu}. The measured value in Eq.~\eqref{eq:Bpitaunu} is more than a factor of 2 bigger -- a  discrepancy with $2.6\sigma$ significance if gaussian errors are assumed (for previous discussion of this tension see ~\cite{Lunghi:2010gv}). In order to avoid having to extrapolate lattice form factor results over the whole $B\to \pi \ell\nu$ phase space, one may also consider only the region of high $q^2 \equiv (p_B-p_\pi)^2$ \cite{Khodjamirian:2011ub},
in which case the discrepancy between the SM expectations and experiment for $\delta \mathcal R^{\pi}_{\tau/\ell} \equiv  \mathcal R^{\pi}_{\tau/\ell} |_{q^2>16\mathrm{GeV}^2}$ is at the level of $1.6\sigma$.

For later convenience we can summarize all the three experimental values as ${\mathcal R^{\pi,\rm exp}_{\tau/\ell}}/{\mathcal R^{\pi,\rm SM}_{\tau/\ell}} =2.38\pm0.66$ (${\delta \mathcal R^{\pi,\rm exp}_{\tau/\ell}}/{\delta\mathcal R^{\pi,\rm SM}_{\tau/\ell}} =1.52\pm0.32$), ${\mathcal R^{\rm exp}_{\tau/\ell}}/{ \mathcal R^{\rm SM}_{\tau/\ell} }=1.49\pm0.26$ and ${\mathcal R^{*,\rm exp}_{\tau/\ell}}/{\mathcal R^{*,\rm SM}_{\tau/\ell}}=1.32\pm0.12$, giving a combined excess of $3.9\sigma$ ($3.4\sigma$) above the SM expectations.
These hints of LFU violations in semileptonic $b\to c$ and $b\to u$ transitions can be contrasted to the pion and kaon sectors where LFU for all three lepton generations has been tested at the percent level and found in excellent agreement with the SM expectations~\cite{LFUPK}\,. 
%(we note in passing that there is a $2\sigma$ tension in $W\to \tau \nu$ decays at LEP, for a recent analysis see \cite{Filipuzzi:2012mg}).

In this Letter we explore the possibility that the hints of LFU violations in semileptonic $B$ decays are due to new physics (NP). We first perform a model independent analysis using effective field theory (EFT), which then allows us to identify viable NP models. Implications for other flavor observables and LHC searches are also derived. 

%%%%%%%%%%%%%%%%%%%%%%%%%%%%%%
\mysection{LFU Violations in $B$ decays and NP}
%%%%%%%%%%%%%%%%%%%%%%%%%%%%%%
We first study  NP effects in $\mathcal R^{(*)}_{\tau/\ell}$ and  $\mathcal R^{\pi}_{\tau/\ell}$ using EFT.  The SM Lagrangian is supplemented with a set of higher dimensional operators, $\mathcal Q_i$, that are generated at a NP scale $\Lambda$ above the electroweak symmetry breaking scale $v= (\sqrt{2}/4G_F)^{1/2}\simeq 174$~GeV
\beq
\mathcal L = \mathcal L_{\rm SM} +  \sum_a \frac{z_a}{\Lambda^{d_a-4}} \mathcal Q_i + \rm h.c.\,,
\label{eq:Lagr}
\eeq
where $d_a$ are the canonical dimensions of the operators $\mathcal Q_a$, and $z_a$ are the dimensionless Wilson coefficients (below we will mostly use rescaled versions $c_a= z_a (v/\Lambda)^{d_a-4}$). We also make two simplifying requirements that at the tree level (i) no dangerous down-type flavor changing neutral currents (FCNCs) and (ii) no LFU violations in the pion and kaon sectors are generated.
 The lowest dimensional operators that can modify 
$R^{(*)}_{\tau/\ell}$ and $\mathcal R^{\pi}_{\tau/\ell}$ then have  the following form, 
\begin{align}
\mathcal Q_{L} &= (\bar q_3 \gamma_\mu \tau^a q_3) \mathcal J^\mu_{3,a}\,, \label{eq:QL}\\
\mathcal Q^{i}_{R} &=  (\bar u_{R,i} \gamma_\mu  b_R) (H^\dagger \tau^a \tilde H)  \mathcal J^\mu_{3,a}\,\\
\mathcal Q_{LR} &=  i \partial_\mu (\bar q_3 \tau^a H  b_R) {\textstyle \sum_j} \mathcal J^\mu_{j,a}\,, \label{QLR} \\
\mathcal Q^{i}_{RL} &=  i \partial_\mu (\bar u_{R,i} \tilde H^\dagger \tau^a q_3)  {\textstyle \sum_j} \mathcal J^\mu_{j,a}\,, \label{eq:QRL}
\end{align}
where $\tau_a = \sigma_a/2$,   $\mathcal J^\mu_{j,a} =  (\bar l_j \gamma^\mu \tau_a l_j)$, $\tilde H \equiv i \sigma_2 H^*$ and $i,j$ are generational indices. 
We work in the down quark mass basis, $q_i = (V^{ji*}_{CKM} u_{L,j},d_{L,i})^T$, and charged lepton mass basis, $l_i = (V^{ji*}_{PMNS}\nu_{L,j},e_{L,i})^T$. Our requirement that there are no down-type tree-level FCNCs means that we impose flavor alignment in the down sector for operators $\mathcal Q_L, \mathcal Q_{LR}$ and $\mathcal Q_{RL}^i$. 
In this way we get rid of {\it all} tree level FCNCs  due to $\mathcal Q_{LR}$ while $\mathcal Q_L$ and $\mathcal Q_{RL}^i$ still generate effects in $c\to u \nu \bar \nu$ and $t \to c(u) \nu\bar\nu$ transitions. The first process is typically obscured by SM tree level contributions (i.e. $D \to (\tau \to \pi\nu) \bar \nu$~\cite{Kamenik:2009kc}), while the second will induce an interesting monotop signature at the LHC~\cite{Kamenik:2011nb}.

Other $d_i\leq  8$ operators can either be reduced to the above using equations of motion, or have vanishing $\bra{0} \mathcal Q_i \ket{B}$ hadronic matrix elements  and thus cannot affect $\mathcal R^{\pi}_{\tau/\ell}$ (e.g., $\bar Q_i \sigma_{\mu\nu} \tau^a H b_R$). Note that $\mathcal Q^i_{L,R}$ are tau lepton flavor specific, while in the case of $\mathcal Q^i_{RL, LR}$ LFU violations are induced by the helicity suppression of the leptonic current, as can be easily seen by integrating by parts and using equations of motion.

In addition, new light invisible fermions $\psi$ could mimic the missing energy signature of SM neutrinos in the $b\to u_i \tau \nu$ decays~\cite{Kamenik:2011vy}. We thus also consider the lowest dimensional operators coupling $\psi$ to SM quarks and charged leptons and invariant under the SM gauge group
\beq\label{eq:Qpsi}
\mathcal Q^{i}_{\psi S} = (\bar q_{i} b_R) (\bar l_3 \psi_R)\,, ~ \mathcal Q^{i}_{\psi V} = (\bar u_{i} \gamma_\mu b_R) (\bar \tau_R \gamma^\mu \psi_R)\,.
\eeq
In the following we consider a single NP operator contributing to $\mathcal R^{(*)}_{\tau/\ell}$ and $\mathcal R^{\pi}_{\tau/\ell}$ at a time and later compare this to some explicit NP model examples.

%%%%%%%%%%%%%%%
\begin{figure}[t]
\centering{
\includegraphics[height=0.23\textwidth]{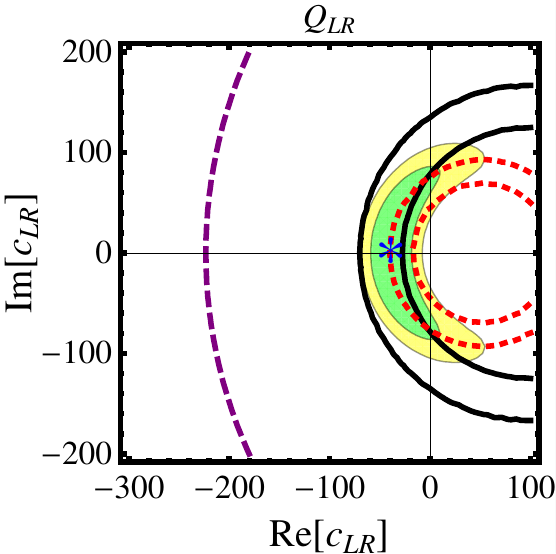}
\includegraphics[height=0.23\textwidth]{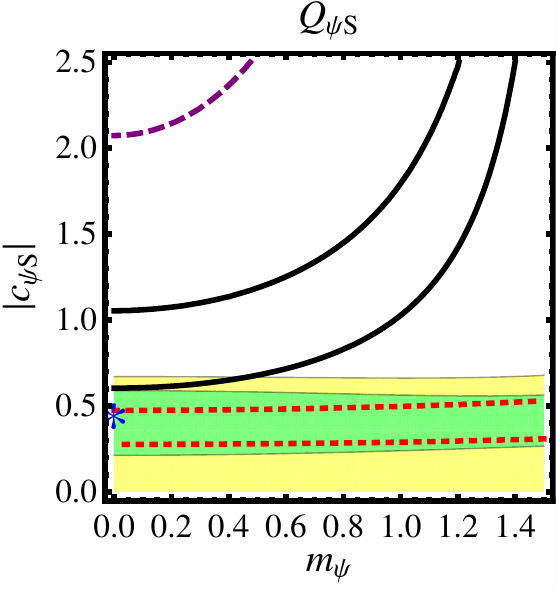}
}
\caption{\footnotesize  Preferred $1\sigma$ (darker green) and $2\sigma$ (lighter yellow) parameter regions for effective operators $\mathcal Q_{LR}$ (left plot, dependence on complex $ c_{LR}$ Wilson coefficient) and MFV $\mathcal Q^i_{\psi S}$ (right plot, dependence on $\psi$ mass and modulus of the universal Wilson coefficient, $| c_{\psi S}|$). The $1\sigma$ constraints from $\mathcal R_{\tau/\ell}$, $\mathcal R^*_{\tau/\ell}$ and $\mathcal R^{\pi}_{\tau/\ell}$ are drawn in full black, dashed purple and dotted red contours, respectively. The best fit points are marked with an asterisk. 
\label{fig:1}}
\end{figure}
%%%%%%%%%%%%%%%

%%%%%%%%%%%%%%%%%%%%%%%%%%%%%%
\mysection{%Operators $\mathcal Q_L$ and $\mathcal Q_{LR}$ and 
Minimal flavor violation}
%%%%%%%%%%%%%%%%%%%%%%%%%%%%%%
The flavor structure of $\mathcal Q_L$ and $\mathcal Q_{LR}$ is completely determined by our requirement that there are no tree level FCNCs in the down sector. The charged currents are then proportional to the same CKM elements as in the SM realizing the Minimal Flavor Violation (MFV) structure~\cite{MFV}.
The effect of  $\mathcal Q_L$ is to rescale the SM predictions for $\mathcal R^{(*)}_{\tau/\ell}$, $\mathcal R^{\pi}_{\tau/\ell}$ by a universal factor $|1+c_L/2|^2$, where $c_L = z_L (v/\Lambda)^2$.   The best fit to the three LFU ratios is obtained for $|1+c_{L}/2| \simeq 1.18$ with a value of $\chi^2 \simeq 4.0$ (for the SM, $\chi^2 \simeq 21$). Both $\mathcal R^{}_{\tau/\ell}$ and $\mathcal R^{*}_{\tau/\ell}$ are then well accommodated, while the $\mathcal R^{\pi}_{\tau/\ell}$  tension remains at the $1.5\sigma$ level. The effective NP scale probed is $\Lambda |z_L|^{-1/2}=v |c_L|^{-1/2} \simeq 0.29$~TeV.

The contributions of $\mathcal Q_{LR}$ can be readily computed using results of~\cite{Kamenik:2008tj, BDTSM}. 
In the case of $\mathcal R_{\tau/\ell}$ we also need to take into account a significant experimental efficiency correction due to the different kinematics induced by the $\mathcal Q_{LR}$ operator compared to $\mathcal Q_{L}$ and the SM~\cite{:2012xj}. Effectively this amounts to multiplying the  term quadratic in $c_{LR}$ by a correction factor of $\sim 1.5$ 
. The same argument applies for the operators $\mathcal Q^i_{RL}$ and $\mathcal Q_{\psi S}$ (near $m_\psi= 0$).
Switching on only the $\mathcal Q_{LR}$ operator the best fit point is  $c_{LR} = z_{LR} (v/\Lambda)^4 \simeq -34$, where $\chi^2 \simeq 8.5$  with both $\mathcal R_{\tau/\ell}$ and  $\mathcal R^{\pi}_{\tau/\ell}$ perfectly accommodated, while a tension with the observed value of $\mathcal R^*_{\tau /\ell}$ remains (see  Fig.~\ref{fig:1} left). Irrespective of $\mathcal R^{\pi}_{\tau/\ell}$, the central measured values of $\mathcal R^{}_{\tau/\ell}$ and $\mathcal R^{*}_{\tau/\ell}$
can never be simultaneously obtained using only $\mathcal Q_{LR}$~\cite{BDTSM}. The preferred value of  $c_{LR}$ points to a very low effective NP scale of $ v |c_{LR}|^{-1/4} \simeq 72$~GeV.

The relative strength of semileptonic $b\to c $ and $b\to u$ transitions generated by the $\mathcal Q_R^i, \mathcal Q_{RL}^i$ or $\mathcal Q_{\psi S,V}^i$ operators is fixed only once we explicitly specify the flavor structure.  
For $\mathcal Q^i_R$, $\mathcal Q^i_{RL}$ and $\mathcal Q^i_{\psi V}$, MFV implies $z^i_{R,RL,\psi V} \propto m_{u_i}$ leading to extremely suppressed effects in  $\mathcal R^{\pi}_{\tau/\ell}$. Consequently we do not consider these operators within MFV. On the other hand, in the case of $\mathcal Q^i_{\psi S}$ the MFV hypothesis  is satisfied by taking $z^i_{\psi S} = V^{ib}_{\rm CKM}  c_{\psi S} (\Lambda/ v)^2$. The corrections to $\mathcal R^{(*)}_{\tau/\ell}$ and  $\mathcal R^{\pi}_{\tau/\ell}$ now also depend on the mass $m_\psi$ of the new invisible fermion $\psi$. 
Close to $m_B-m_{D^{(*)}}-m_\tau$ thresholds the contributions to $\mathcal R^{(*)}_{\tau/\ell}$ are suppressed relative to the  ones in $\mathcal R^{\pi}_{\tau/\ell}$. Varying both $c_{\psi S}$ and $m_\psi$ the best fit of $\chi^2=9.1$ is reached for $ c_{\psi S} \simeq 0.40$ and $m_\psi = 0$ (see Fig.~\ref{fig:1} right). 
Significant tensions between the three observables remain.

%%%%%%%%%%%%%%%%%%%%%%%%%%%%%%
\mysection{Generic flavor structures}
%%%%%%%%%%%%%%%%%%%%%%%%%%%%%%
In the presence of more general flavor violation the NP contributions to $\mathcal R^{\pi}_{\tau/\ell}$ are no longer related to those in $\mathcal R^{(*)}_{\tau/\ell}$. 
We thus parametrize the contributions of $\mathcal Q_{\psi S,\psi V}^i$ and $\mathcal Q_{R,RL}^i$ operators to $b\to c$ semileptonic transitions by ($A=\psi S,\psi V,R,RL$) $z^c_{A} = c_{A} (\Lambda/ v)^2$,  and to $b\to u$ semileptonic transitions by  $z^u_{A} = \epsilon_{A}  z^c_{A}$. 
The effect of $\mathcal Q^i_{R}$ is to rescale the SM expectations $\mathcal R_{\tau/\ell}$ by $|1- c_{R}/2V_{cb}|^2$ and  $\mathcal R^{\pi}_{\tau/\ell}$ by $|1+ \epsilon_R c_{R}/2V_{ub}|^2$. For  $\mathcal R^*_{\tau/\ell}$ we obtain
$
% \beq
{\mathcal R^{*,R}_{\tau/\ell}}/{\mathcal R^{*,\rm SM}_{\tau/\ell}} =  \left. 1 - 0.88\, {\rm Re}( c_{R}/V_{cb} ) +0.25 | c_{R}/V_{cb} |_.^2  \right.
 %\eeq
$
On the other hand contributions of $\mathcal Q_{RL}^i$ can be obtained from the corresponding expressions for $\mathcal Q_{LR}$ in the previous section  with obvious modifications for the different flavor and chiral structure.

\begin{figure}[t]
\centering{
\includegraphics[height=0.23\textwidth]{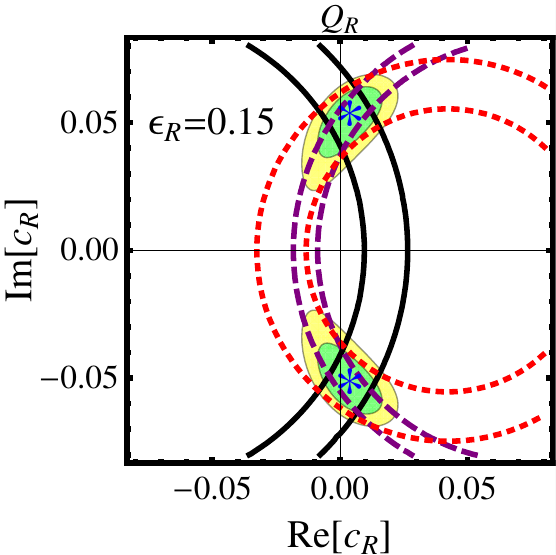}
\includegraphics[height=0.23\textwidth]{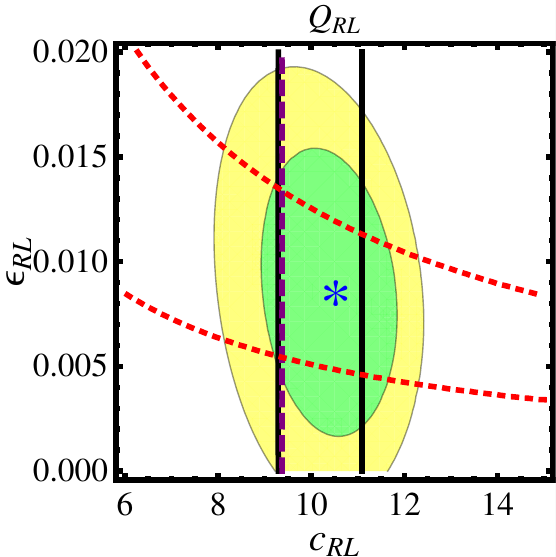}
}
\caption{\footnotesize  Preferred parameter regions for effective operators $\mathcal Q^i_{R}$ (left plot, as a funciton of complex $ c_R$ Wilson coefficient, and $ \epsilon_R$ fixed to the best fit value),  and for $\mathcal Q^i_{RL}$ (right plot, as a function of real $ c_{RL}$ Wilson coefficient and the mixing ratio $ \epsilon_{RL}$). Contours and shaded regions as in Fig.~\ref{fig:1}. The best fit points are marked with an asterisk.}
\label{fig:2}
\end{figure}

We fit the data to pairs of $( c_i, \epsilon_i)$  using CKM inputs from the global fit~\cite{Charles:2011va}. The results are presented in Fig.~\ref{fig:2}.
%%%%%%%%%%%%%%%
%%%%%%%%%%%%%%%
Considering the NP operator $\mathcal Q^i_{R}$,  all three LFU ratios can now be perfectly accommodated at $ c_R \simeq -0.0039 \pm 0.053 i$ and $ \epsilon_R \simeq 0.15$ (analogously for $\mathcal Q^i_{\psi V}$ at $m_\psi =0$, $|c_{\psi V}| \simeq 0.050$ and $\epsilon_{\psi V} \simeq 0.15$). Interestingly, a good fit necessarily implies the presence of large CP violating contributions, suppressed by an effective NP scale $v |\rm Im( c_R)|^{-1/4} \simeq 0.36$~TeV. Similarly, NP contributions from $\mathcal Q^i_{RL}$  can now simultaneously explain both $\mathcal R^{\pi}_{\tau/\ell}$ and  $\mathcal R^{(*)}_{\tau/\ell}$ discrepancies. The best fit of $\chi^2\simeq 0.54$ is obtained at $ c_{RL}\simeq  11$ and $ \epsilon_{RL} \simeq 0.0084$. The required size of NP contributions points to a low NP scale of $v |c_{RL}|^{-1/4} \simeq 97$~GeV. On the other hand,  generic flavor structures do not significantly improve the MFV fit of the $\mathcal Q^i_{\psi S}$ operator, due to the tension between $\mathcal R_{\tau/\ell}$ and $\mathcal R^*_{\tau/\ell}$  (present also for $\mathcal Q_{LR}$, see Fig.~\ref{fig:1}). Nonetheless, both $\mathcal R_{\tau/\ell}$ and $\mathcal R^{\pi}_{\tau/\ell}$ can now be accommodated simultaneously provided the parameters are near $m_\psi = 0$, $ c_{\psi S} \simeq 0.04$ and $ \epsilon_{\psi S} \simeq 0.03$ (at which point $\chi^2 \simeq 5.4$)\,. 

%%%%%%%%%%%%%%%%%%%%%%%%%%%%%%
\mysection{Explicit models}
%%%%%%%%%%%%%%%%%%%%%%%%%%%%%%
Specific NP models in general contribute to more than one operator of the effective Lagrangian in Eq.~\eqref{eq:Lagr}. The agreement with data  for $\mathcal Q_{RL}^i$ operators suggests an obvious candidate -- the two-Higgs doublet model (2HDM), where charged Higgs ($H^+$) exchanges generate both $c_{RL}$ and $c_{LR}$.  No tree level FCNCs arise in 2HDMs with natural flavor conservation where the two Higgs doublets couple exclusively to $u_R$ and/or $d_R$. The four types of natural flavor conservation 2HDMs: Type I, Type II, ``lepton specific" and ``flipped"~\cite{Branco:2011iw} give $c_{LR} = (2 m_b v / m_{H^+}^2) \{ {\rm ctg}^2\beta, {\rm tg}^2\beta, -1, -1  \}$ and $c^i_{RL} = (2 m_{u}^i v / m_{H^+}^2) \{ {\rm ctg}^2\beta, -1, -1, {\rm ctg}^2\beta  \}$, respectively. Here ${\rm tg}\beta$ is the ratio of the two Higgs doublets' vacuum expectation values. Imposing the $m_{H^+} \gtrsim 80$~GeV bound from direct searches at LEP~\cite{Heister:2002ev} (even stronger constraints can been derived from one-loop charged Higgs contributions to FCNC processes~\cite{Deschamps:2009rh}) and $\mathcal O(1) \lesssim {\rm tg} \beta \lesssim \mathcal O (100)$ so that the Yukawas are perturbative, we find that none of the natural flavor conservation 2HDMs can simultaneously account for the three LFU ratios. 

In principle there is enough freedom in the Higgs couplings to quarks to explain the observed LFU ratios using 2HDMs with more general flavor structure. A simple limit is that only one of the Higgs doublets obtains a vacuum expectation value. The  charged Higgs is then part of the remaining Higgs doublet ($\bar H$). The interaction terms ${\cal L} \supset \kappa_{RL}^i\bar q_3 u_R^i \bar H + \kappa_{LR}^i\bar b_R \bar H^\dagger q_i + \kappa^\tau \bar \tau_R l_3 \bar H +\rm h.c.$ generate $c_{RL}^{i\tau}=- \kappa_{RL}^{i*} (\kappa^\tau v/m_\tau) (v/m_{H^+})^2$ and $c_{LR}^{i\tau}=- \kappa_{LR}^{i*} (\kappa^\tau v/m_\tau) (v/m_{H^+})^2$ Wilson coefficients for the $i \partial_\mu (\bar u_i  \tilde H^\dagger \tau^a   q_3) \mathcal J^\mu_{3,a}$  and $i \partial_\mu (\bar q_i \tau^a H  b_R) \mathcal J^\mu_{3,a}$ operators, generalizations of Eqs.~\eqref{eq:QRL} and~\eqref{QLR}, respectively. The best fit regions have a fourfold amiguity with two solutions for $(\kappa_{LR}^u-\kappa_{RL}^u)\kappa^\tau\simeq \{0.9,-4\}\cdot10^{-3} (m_{H^+}/v)^2$, and two solutions for $(\kappa_{RL}^c\kappa^\tau,\kappa_{LR}^c\kappa^\tau)\simeq \{(-6,8),(-12,2)\}\cdot 10^{-2}(m_{H^+}/v)^2$.  These values are large enough to pose severe flavor building problems. The products $\kappa_{RL}^{c(u)} \kappa^\tau$  are roughly three (four) orders of magnitude larger than the corresponding Yukawas giving fermion masses, $(m_{c(u)}/v)(m_\tau/v)$. 
Furthermore, in order to satisfy FCNC bounds from $D^0$, $B_s$ and $B_d$ mixing, there needs to be at least an order of magnitude cancellation between different  contributions even for $\kappa^\tau=1$ (to suppress $\Delta B=2$ transitions, a viable solution is also $\kappa_{LR}^i=0$). 
If such a charged Higgs is lighter than the top quark, it could be observed in $t\to b H^+$ decays. The null results of existing searches at ATLAS and CMS imply $|\kappa^t_{RL,LR}| \lesssim {\mathcal O}(0.2-0.4)$ for the $H^+$ mass between $80$ GeV and $160$ GeV \cite{Aad:2012tj}. If the charged Higgs is heavier than the top, the dominant signal could come from $g b\to H^- t$ production with, e.g., the $pp$ cross section at the 8~TeV LHC of $1.4{\rm pb} (|\kappa_{RL}^t|^2+|\kappa_{LR}^t|^2)$ for $m_{H^-}=200$ GeV. Also for larger $H^+$ masses $\tau+\slashed E_T$ and $tb$ resonance searches~\cite{Aad:2012ej} become effective, since $H^-$ then decays predominantly to $\bar t b$ and $\tau \nu$ depending on the relative sizes of $\kappa_{LR,RL}^t$ and $\kappa^\tau$. 

An alternative possibility is represented by leptoquarks. In particular, scalar leptoquarks forming the $({\bf 3},{\bf 3},-1/3)$, $({\bf\bar 3},{\bf 2}, -7/6)$ and $({\bf 3}, {\bf 1}, -1/3)$ representations of the SM $SU(3)_c\times SU(2)_L\times U(1)_Y$ gauge group as well as vector leptoquarks in the $({\bf 3},{\bf 3},2/3)$, $({\bf\bar 3},{\bf 2}, 5/6)$ and $({\bf 3}, {\bf 1}, 2/3)$ representations can contribute to (semi)leptonic charged current meson decays at the tree level. In general they will also induce dangerous FCNC operators and are thus potentially severely constrained~\cite{Dorsner:2009cu}. As an example  we consider the scalar electroweak triplet leptoquark $S_3 = ({\bf 3},{\bf 3},-1/3)$ with renormalizable interactions to 3rd generation SM fermions (aligned with the mass basis of down-like quarks and charged leptons)
$
\mathcal L^{}_{} \supset Y_{S_3}  \overline{q_3^c} i \sigma_2 \tau^a S_3^{a*} l_3 + \rm h.c.\,. 
%\label{eq:LQ}
$
Integrating out $S_3$ at the tree level induces a contribution to $\mathcal Q_L$ with $c_L = (|Y_{S_3}|^2 / 4) (v/m_{S_3})^2$. LFU violations in $B\to D^{(*)}$ transitions (and partially $\mathcal R^\pi_{\tau/\ell}$) can then be accommodated provided $|Y_{S_3}|/m_{S_3} \simeq 1/150$~GeV. The most severe constraints on these parameters come from electroweak precision tests~\cite{ewpt} requiring $|Y_{S_3}|/m_{S_3} \lesssim 1/450$~GeV, in tension with the value preferred by  $B$ decays.  
Additional contributions to electroweak precision observables from the UV completion of the effective $S_3$ model could soften this tension.
Most constraining direct bound on the mass of $S_3$ is from the CMS search for 3rd generation scalar leptoquarks decaying to $b \nu$~\cite{CMS}. Taking into account the $S_3\to b\nu$ branching ratio we obtain a bound $m_{S_3} \gtrsim 280$~GeV. Future dedicated searches using also the $t \tau$ decay channel~\cite{Davidson:2011zn} or associated $S_3$ production with the monotop signature~\cite{Kamenik:2011nb} could further constrain this model. 

Modifications of semileptonic transitions involving the third generation quarks and leptons are also expected in models of strong electroweak symmetry breaking or composite Higgs models where the heavier SM fermions are expected to be partially or mostly composite~\cite{Kaplan:1991dc}. The exchange of strong sector vector resonances will induce contributions to $\mathcal Q_{L,R}$ which can be parametrized as
\beq
\frac{z_L}{\Lambda^2} \sim  \frac{g_\rho^2}{m_\rho^2} [f^q_3]^2  [f^l_3]^2\,, ~~ \frac{z^{u(c)}_R}{\Lambda^4} \sim  \frac{g_\rho^2 }{ m_\rho^2 } \frac{y^{Qd}_3 y^{Qu(c)}_{1(2)}}{m_Q^2} [f^l_3]^2\,, 
\eeq 
where $g_\rho\lesssim \sqrt{4\pi}$ and $m_{\rho}\sim \mathcal O({\rm TeV})$ are the strong sector vector resonance coupling and mass, while $m_{Q} \lesssim  \mathcal O({\rm TeV})$ is the mass of  the strong sector fermion resonances ($Q$) transforming as $({\bf 3},{\bf 2},1/6)$ under the SM gauge group. Furthermore, $f_i^{q,l} \in [0,1]$ are compositeness fractions of $i$-th generation left handed quarks and leptons respectively, parametrizing the mixing of chiral fermions with strong sector fermion resonances (again assuming down type mass alignment), while $y^{Qd,Qu}_i$ are the couplings of right-handed chiral up- and down-type quarks to the composite Higgs and $Q$ fermion resonance fields in $\mathcal L \supset  y^{Qd}_i \bar Q H d^i_R + y^{Qu}_i \bar Q \tilde H u^i_R + \rm h.c.$. For concreteness we fix $f_3^l=f_3^q=1$ (third generation compositeness, the first two generations of left-handed fermions can be completely elementary), $g_\rho=\sqrt{4\pi}$ and fit $m_\rho$, $\epsilon_{32} \equiv y^{Qd}_3 y^{Qu}_2 v^2/m_Q^2$ and $\epsilon_{31} \equiv y^{Qd}_3 y^{Qu}_1 v^2/m_Q^2$ to the three LFU ratios.  A good fit to all three observables is obtained for $m_\rho \simeq 1$~TeV in two regions, around $\epsilon_{32}\simeq 0$ and $\epsilon_{31} \simeq -0.006$ but also $\epsilon_{32}\simeq 0.009$ and $\epsilon_{31} \simeq 0.04$. Note that non-zero $\epsilon_i$ (signaling compositeness of right-handed quarks) are required to fit $\mathcal R^{(*)}_{\tau/\ell}$ and $\mathcal R^\pi_{\tau/\ell}$ simultaneously within $1\sigma$.
Similarly to the $H^+$ in 2HDMs, the strong sector charged vector resonances are susceptible to $\tau+\slashed E_T$ and $tb$ resonance searches~\cite{Aad:2012ej} at the LHC. Another interesting channel is the resonant or Higgs associated production of fermionic $Q$ resonances through $d_i g$ or $u_i g$ fusion.

\mysection{Prospects}
It is important that the indications of LFU violation in $B\to D^{(*)}\tau\nu$ and $B\to \tau\nu$ decays are verified using related $B$ decays. Measuring the $ B \to \pi \tau  \nu$ branching ratio~\cite{Khodjamirian:2011ub} could confirm LFU violation in $b\to u\tau \bar \nu$ transitions. In the SM one has
\beq
\left[{\mathcal B(B\to \pi \tau\nu)}/{\mathcal B(B\to \pi \ell\nu)}\right]^{\rm SM} = 0.68\pm 0.03\,,
\eeq
using form factor estimates from the lattice~\cite{Laiho:2009eu}, and where the uncertainty is dominated by the shape of the scalar $B\to \pi$ form factor as extracted from the lattice simulation~\cite{Dalgic:2006dt}.
A measurement of this observable should be possible at the (super)B-factories. It would also help to disentangle the possible underlying NP, since different effective operators in Eqs.~\eqref{eq:QL}--\eqref{eq:QRL} give different contributions compared to $\mathcal R_{\tau/\ell}^\pi$. A similarly useful observable probing LFU in $b\to c$ transitions would be the purely leptonic decay of the $B_c$ meson $B_c\to \tau\nu$. The experimental prospects for such a measurement are more uncertain, however. 

The NP interpretations of the LFU violation in $B$ decays have also interesting implications for the direct searches at the LHC. 
%The  details of the relevant LHC signatures are model dependent. 
%The relevant results of the model specific  searches for charged Higgs, leptoquarks and strong sector resonant states that are already being performed by CMS and ATLAS have been briefly reviewed above. 
In addition to the model dependent searches for on-shell production of the relevant NP states,
there are also some generic signatures that are more tightly related to the fact that LFU violation is seen in $B$ decays. All the models either predict contributions to $h+\tau+ \slashed E_T$ channel where $h$ is the physical neutral Higgs boson (for the models that match onto $\mathcal Q_R^i$, $\mathcal Q_{LR}$ and $\mathcal Q_{RL}^i$ EFT operators), the monotop $t+\slashed E_T$ signature (for $\mathcal Q_L$ and $\mathcal Q^i_{RL}$ operators)
 or  the $(t+)\tau+ \slashed E_T$ channels. The latter are possible for all EFT operators, but the final state with a top quark is directly related to the strength of $B$ decay LFU anomalies only for $\mathcal Q_L$ and $\mathcal Q_{LR}$ operators. 
 %For instance in the 2HDM the $h \tau\nu$ final state would arise from resonant production $pp\to H^-\to h H^-$, however this is suppressed compared to $g b\to H^- t$ by the smallness of $\kappa^u_{RL}$.

In conclusion, we have shown that the indications for the violation of LFU in $B$ decays can only partly be explained in presence MFV modifications of the left-left vector currents coupling to the third generation quarks and leptons. A better fit to current data is provided by non-MFV right-right vector or right-left scalar currents. We have also shown how such effects could arise from  2HDMs, from leptoquark models or from models with composite quarks and leptons.

\begin{acknowledgments}

We thank D. Be\v cirevi\'c, A. Falkowski, and R. Harnik for useful discussions. This work was supported in part  by the Slovenian Research Agency. JZ was supported in part by the U.S. National Science Foundation under CAREER Award PHY-1151392.

\end{acknowledgments}

\end{document}